\documentclass{WileyMSP-template}
\usepackage{amsmath}
\usepackage[dvipsnames]{xcolor}
\begin{document}

\pagestyle{fancy}
\rhead{\includegraphics[width=2.5cm]{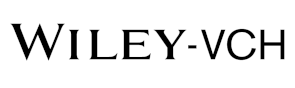}}

\title{Triclinic metamaterials by tristable origami with reprogrammable frustration}

\maketitle


\author{Ke Liu}
\author{Phanisri P. Pratapa}\footnote{K.L. and P.P.P. contributed equally to this work.}
\author{Diego Misseroni}
\author{Tomohiro Tachi}
\author{Glaucio H. Paulino*}



\begin{affiliations}
K. Liu\\
Department of Advanced Manufacturing and Robotics, Peking University, Beijing 100871, China\\

P. P. Pratapa\\
Department of Civil Engineering, Indian Institute of Technology Madras, Chennai 600036, TN, India\\

D. Misseroni\\
Department of Civil, Environmental and Mechanical Engineering, University of Trento, Trento 38123, Italy\\

T. Tachi\\
Graduate School of Arts and Sciences, University of Tokyo, Tokyo 153-8902, Japan\\

G. H. Paulino\\
Department of Civil and Environmental Engineering, Princeton University, Princeton, New Jersey, 08544, USA\\
Princeton Institute for the Science and Technology of Materials, Princeton University, Princeton, New Jersey, 08544, USA\\
Email Address: gpaulino@princeton.edu

\end{affiliations}



\begin{abstract}

Geometrical frustration induced anisotropy and inhomogeneity are explored to achieve unique properties of metamaterials that set them apart from conventional materials. According to Neumann's principle, to achieve anisotropic responses, the material unit cell should possess less symmetry. Based on such guidelines, we present a triclinic metamaterial system of minimal symmetry, which is originated from a Trimorph origami pattern with a simple and insightful geometry: a basic unit cell with four tilted panels and four corresponding creases. The intrinsic geometry of the Trimorph origami, with its changing tilting angles, dictates a folding motion that varies the primitive vectors of the unit cell, couples the shear and normal strains of its extrinsic bulk, and leads to an unusual Poisson's effect. Such effect, associated to reversible auxeticity in the changing triclinic frame, is observed experimentally, and predicted theoretically by elegant math formulae. The nonlinearities of the folding motions allow the unit cell to display three robust stable states, connected through snapping instabilities. 
%
%
When the tristable unit cells are tessellated, phenomena that resembles linear and point defects emerge as a result of geometric frustration. The frustration is reprogrammable into distinct stable and inhomogeneous states by arbitrarily selecting the location of a single or multiple point defects. The Trimorph origami demonstrates the possibility of creating origami metamaterials with symmetries that were hitherto non-existent, leading to triclinic metamaterials with tunable anisotropy for potential applications such as wave propagation control and compliant micro-robots.

\end{abstract}

\section{Introduction}
Anisotropic materials have properties that vary with respect to different spatial directions. Such feature is preferred in many applications, for instance, when the intended use is to carry loading that requires different stiffness and strength in different directions. While many materials might have the required stiffness and strength, an anisotropic material might display higher strength to weight ratio along preferential directions \cite{Cowin2007_tissue}. Bone tissue, wood, nacre, and muscles, are all examples of anisotropic materials \cite{Cowin2007_tissue,song_processing_2018,gao_materials_2003}. These examples from nature follow the Neumann's principle~\cite{newnham2005properties}, which states that to achieve anisotropic responses, the material microstructures should possess less geometric symmetry. Following advances in manufacturing, researchers have been able to mimic natural materials by creating mechanical metamaterials with engineered subscale microstructures, offering a variety of special and unusual properties~\cite{schwaiger_extreme_2019,sanders_optimal_2021,injeti_metamaterials_2019,Haghpanah2016,fleck_micro-architectured_2010,zhang_hierarchical_2021,meza_strong_2014,hussein_dynamics_2014}, such as negative thermal expansion \cite{wang_lightweight_2016, cabras2019micro}, negative Poisson’s ratio \cite{Bertoldi2017,morvaridi2021hierarchical,PradeepLiu2019}, vanishing shear modulus \cite{kadic_practicability_2012}, and shear-normal coupling \cite{frenzel_three-dimensional_2017,lipton_handedness_2018}. However, most existing designs of mechanical metamaterials have properties that are limited to either isotropic or orthotropic symmetries. The full spectrum of anisotropic responses is yet to be explored. For instance, it is unclear how common properties defined under isotropic or orthotropic symmetry could be generalized, and how traditionally independent properties would couple with each other, in systems with less or zero symmetries.

Among different types of unit cell symmetries for a periodic system, the triclinic symmetry is the one that yields fully anisotropic properties \cite{Cowin2007_tissue,zheng_description_1994,podesta_symmetry_2019}. The triclinic symmetry describes a periodic system whose primitive vectors are of unequal length, and the angles between these vectors are all different and may not even include 90$^\circ$. Due to its rich design space, origami structures have been a major source of inspiration for creating metamaterial microstructures with various symmetry types \cite{Filipov2015a,Zhai2018,Mukhopadhyay2020,pratapa_reprogrammable_2021,liu_bio-inspired_2021,sadoc_geometrical_1999,Waitukaitis2015,Miyazawa2021}. In the literature, some tubular origami-based metamaterials have been created to achieve triclinic symmetry \cite{Overvelde2017}. However, the disadvantage of tubular designs is that their unit cell geometry and configuration space are typically intricate, involving several parameters. Consequently, the energy landscapes of their tessellations are usually difficult to program \cite{IniguezRabago2019}, which is critical for generating reprogrammability. Conversely, in this work, we introduce a simple and effective origami pattern composed of degree-4 unit cells (consisting of four tilted panels and four corresponding creases), which is assembled into a class of triclinic mechanical metamaterials displaying reprogrammable defects, with neither rotational nor reflective symmetry. 

\begin{figure}[!ht]
	\centering
	\includegraphics[width=0.8\linewidth]{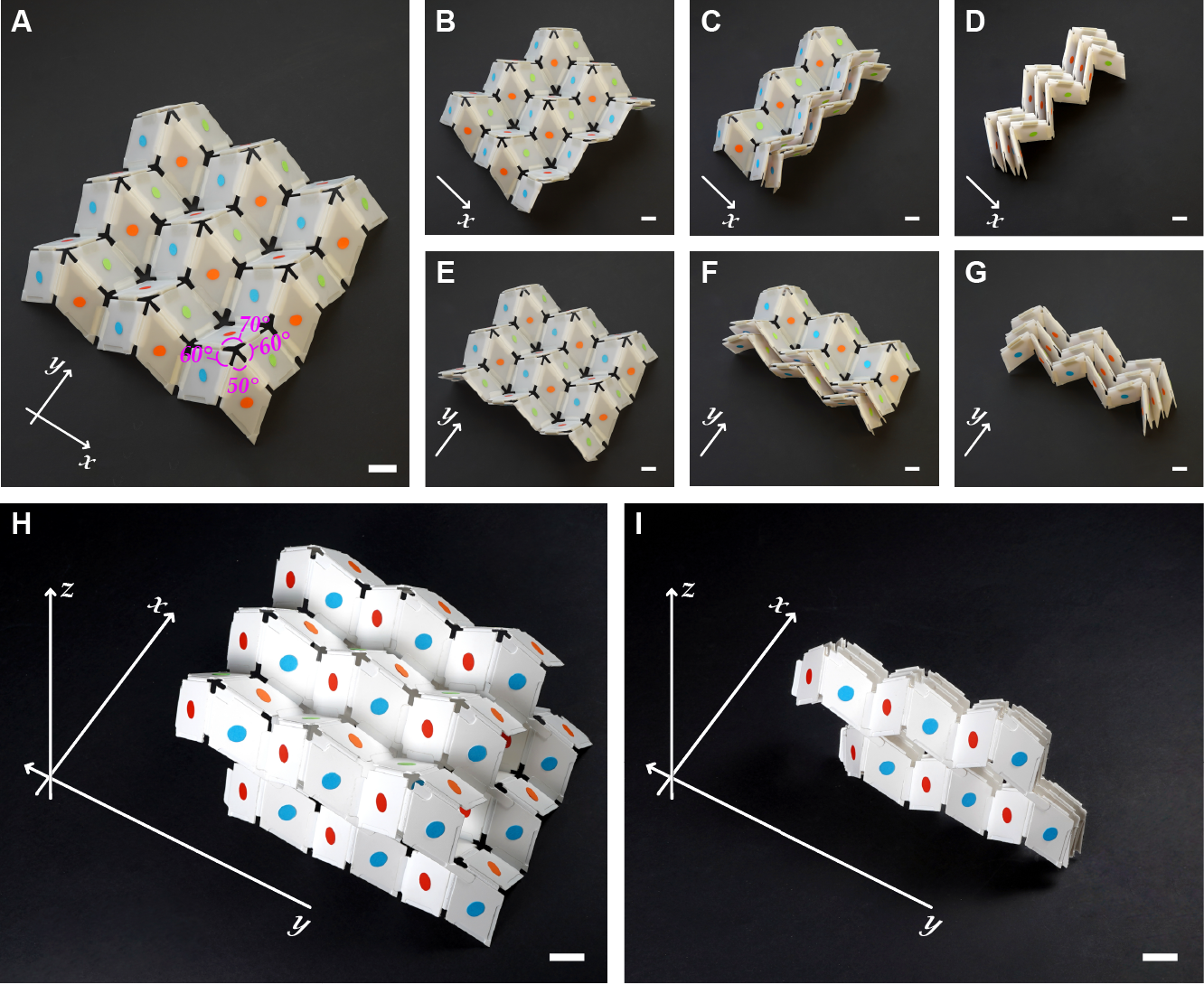}
	\caption{Trimorph origami-based triclinic metamaterials. (\textbf{A}) A piece of metamaterial based on 2D tessellation of the Trimorph origami. (\textbf{B-G}) Different self-stressed stable configurations of the metamaterial shown in A. (\textbf{H}) A 3D metamaterial assembly obtained by stacking the 2D metamaterial. (\textbf{I}) A different stable configuration of the 3D metamaterial, in analogy to state (\textbf{D}) of the 2D metamaterial. Scale bar: 20mm.}
	\label{fig:intro}
\end{figure}

The aforementioned origami, named the Trimorph pattern, can be continuously folded into three distinct modes along the kinematic path and two flat-folded states, allowing the metamaterial unit cell to reconfigure itself and hence significantly change all the Bravais lattice parameters of the triclinic crystal family (three angles and three lengths).
Consequently, the elastic properties of the metamaterial are tunably anisotropic, leading to unusual Poisson's effect and shear-normal coupling in the changing triclinic frame. By tuning the fold energy parameters, we can show that the unit cell has three stable states, each residing in a different mode. Zooming out from the unit cell to 1D, 2D, and 3D assemblies, we show that the resultant metamaterial can switch reversibly among different frustrated states, causing an initially homogeneous system to have intended inhomogeneity, as shown in Fig.~\ref{fig:intro}. As the first report of this triclinic metamaterial, we would mainly focus on the behavior of the Trimorph unit cell and resulting 2D tessellations. However, 3D assemblies are possible by stacking the 2D tessellations, as shown in Figs.~\ref{fig:intro}H and~\ref{fig:intro}I, whose mechanical behavior is largely inherited from their 2D parents.

In summary, we investigate the Trimorph pattern through mathematical analyses, numerical simulations, and experimental validation, including both rigid and non-rigid behaviors. We propose a theory to quantify the Poisson's effect in the changing triclinic frame through the lattice Poisson's ratio. To quantify the unusual Poisson's effect experimentally, we establish both a manufacturing technique for this non-developable pattern, and an experimental device named the \textit{Saint-Venant setup}.
According to the Saint-Venant principle \cite{Timoshenko1951}, extra zones near the boundary of a tested sample must be excluded when evaluating the properties of the material, which leads to a need for large enough samples in conventional mechanical testing to ensure a uniform deformation in the central portion of the sample.
We demonstrate that the \textit{Saint-Venant setup} alleviates the influence of unwanted boundary effects, leading to precise and reliable measurements on relatively small samples that represent the physics of the parent periodic system~\cite{misseroni2022experimental}. We further observe that the Trimorph metamaterial displays equal but opposite Poisson's ratio under stretching and bending by our generalized lattice-based definition (this was previously observed in standard origami metamaterials only when their lattice and principal Poisson's ratios coincide, i.e. under strict orthotropic symmetry conditions~\cite{PradeepLiu2019, Schenk2013a, Wei2013}).  
We discover the existence of line and point defects in the multistable Trimorph based metamaterial, and study their scaling effect, which is relevant for actual applications. We identify that the point defect causes significant frustration of the metamaterial. As both the line and point defects are recoverable, we can control the location of the defects in a piece of metamaterial, and thus reprogram its frustrated state(s). As the aforementioned manufacturing technique allows precise control of the properties of each folding hinge, we are able to observe and demonstrate the defects on physical samples extracted from periodic systems.

\section{Triclinic Configuration Space}
\begin{figure}[!ht]
	\centering
	\includegraphics[width=0.8\linewidth]{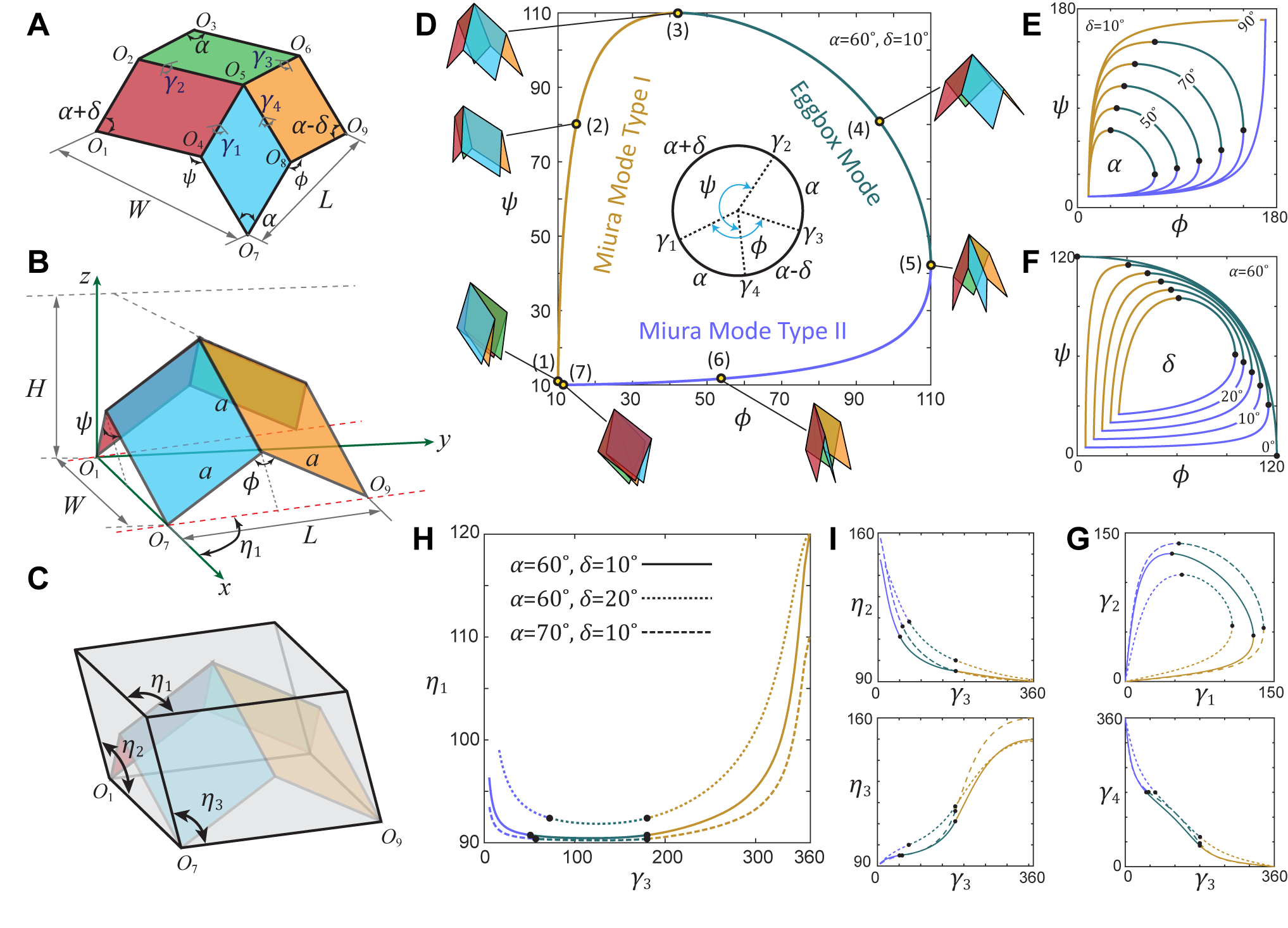}
	\caption{Geometry of the Trimorph unit cell. (\textbf{A}) Schematic with notation of vertices, panel angles, and folding angles. (\textbf{B}) Sketch of a Trimoprh unit cell in the Cartesian frame. (\textbf{C}) The triclinic bounding box of the Trimorph unit cell. (\textbf{D}) The kinematic path that shows all configurations during folding. The colors of the panels in the insets consistently follows the color code in A and B. (\textbf{E-F}) Variations of the kinematic path due to change of the defining angles of the Trimorph pattern, i.e., $\alpha$ and $\delta$. (\textbf{G}) Relationships between the folding angles: $\gamma_1$ vs. $\gamma_2$ and $\gamma_3$ vs. $\gamma_4$. (\textbf{H}) The triclinic lattice angle $\eta_1$ vs. folding angle $\gamma_3$. (\textbf{I}) $\eta_2$ vs. $\gamma_3$ and $\eta_3$ vs. $\gamma_3$.}
	\label{fig:geom}
\end{figure}

To understand the mechanical behaviour of the triclinic metamaterial, we start by examining the geometry of the Trimorph origami. A Trimorph unit cell consists of four rhombus panels, as shown in Figs.~\ref{fig:geom}A and~\ref{fig:geom}B. We denote the vertices as $O_1$ to $O_9$, the folding angles as $\gamma_1$ to $\gamma_4$, and the two angles between opposite creases as $\phi$ and $\psi$. The four panels are characterized by angles $\alpha$, $\delta$, and uniform side length $a$. Compared to the well-known Miura-ori and eggbox patterns \cite{Schenk2013a,Wei2013,nassar2017curvature}, the Trimorph pattern distinguishes itself by having a triclinic symmetry, which means that the bounding box of a Trimorph unit cell is composed of non-orthogonal faces, as shown in Fig.~\ref{fig:geom}C. Taking the parallelogram $O_1 O_7 O_9 O_3$ as a base, if $O_1 O_7$ is placed along the $x$-direction, $O_7 O_9$ is not parallel to the $y$-axis. The folding kinematics of a Trimorph unit cell is described by an implicit function of the opposite crease angles $\phi$ and $\psi$:
\begin{align}\label{eq:fphipsi}
	f(\phi,\psi) = 4\cos^2\phi\cos^2\psi-4(\cos^2\phi+\cos^2\psi) +
	16 \xi_1 (\cos\phi+\cos\psi) - 8 \xi_2 \cos\phi\cos\psi - \xi_3 = 0.
\end{align}
The coefficients are given by:
\begin{align}
	\xi_1 &= \cos^2\alpha\cos\delta, \\
	\xi_2 &= (\cos2\alpha+\cos^2\delta), \\
	\xi_3 &= \sin^22\delta + \cos^2\delta(4+8\cos2\alpha).
\end{align}
Clearly, $f(\phi,\psi) = f(\psi,\phi)$, which reflects an algebraically symmetric role of $\phi$ and $\psi$, as plotted in Fig.~\ref{fig:geom}D. 

\begin{figure}[!ht]
	\centering
	\includegraphics[width=0.75\linewidth]{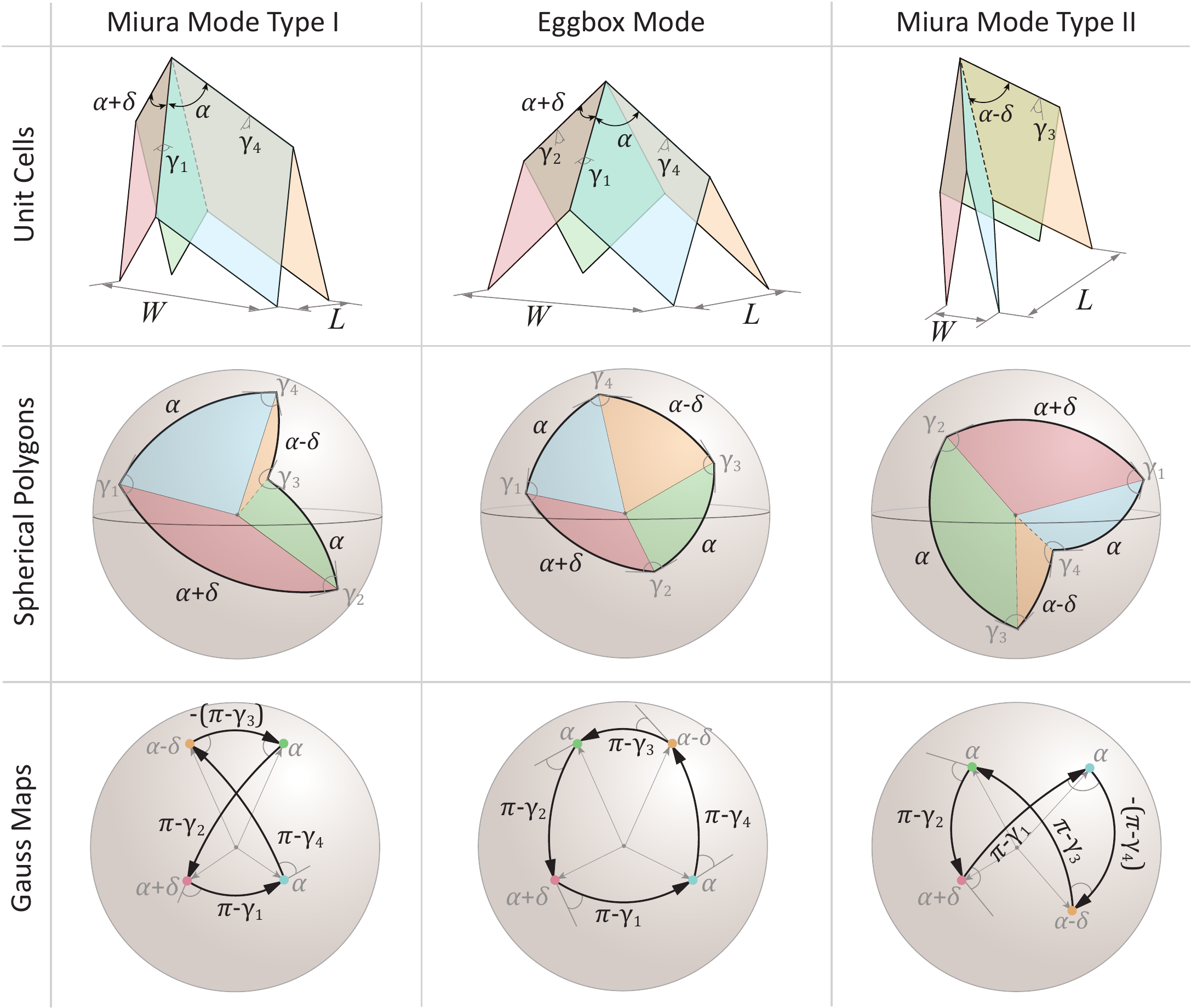}
	\caption{Spherical polygon and Gauss map representations of the three characteristic modes of the Trimorph unit cell. 
	This figure connects the math of spherical trigonometry and the programmable states of matter.}
	\label{fig:table}
\end{figure} 

Different ranges of $\phi$ and $\psi$ lead to three modes of the Trimorph unit cell, which are: Miura mode - type I, eggbox mode, and Miura mode - type II. The eggbox mode has four mountain folds (inset (4) in Fig.~\ref{fig:geom}D). The two Miura modes have three mountain folds and one valley fold, similar to the well-known Miura-ori pattern. The two Miura modes are different as in type I, $O_5 O_6$ is a valley fold with $\pi < \gamma_3 < 2 \pi$ (insets (1), (2) in Fig. \ref{fig:geom}D); while in type II, $O_5 O_8$ is a valley fold with $\pi < \gamma_4 < 2 \pi$ (insets (6), (7) in Fig.~\ref{fig:geom}D; also, see Fig.~\ref{fig:table}). The three modes are topologically different in terms of their Gauss maps, as shown in Fig.~\ref{fig:table}. While the eggbox mode projects a convex spherical quadrilateral, the two Miura modes project spherical bow-ties in two different orientations. The two transition states between the three modes have degenerate creases (either $O_5 O_6$, or $O_5 O_8$) that become flat (insets (3), (5) in Fig.~\ref{fig:geom}D). The Trimorph unit cell has two flat folded states, as shown by the insets (1), (7) in Fig.~\ref{fig:geom}D, with distinct orders of folded panels. Varying the values of design variables $\alpha$ and $\delta$, we obtain different shapes of the implicit function $f(\phi,\psi)$ (Figs.~\ref{fig:geom}E-F). When $\alpha = 90 ^ \circ$, the Trimorph pattern becomes the Barreto Mars pattern \cite{evans_rigidly_nodate} with the eggbox mode vanishing; When $\delta = 0 ^ \circ$, the Trimorph pattern degenerates to the standard eggbox pattern with the two Miura modes vanishing. These are particular cases obtained from the intrinsic geometric parametrization of the pattern.  

The folding angles can be derived using spherical trigonometry from $\phi$ and $\psi$ (Supporting Information). Their mutual relationships are plotted in Fig.~\ref{fig:geom}G. To describe the folding kinematics of a Trimorph unit cell, both $\phi$ and $\psi$ are needed, because only using either one of the two leads to ambiguous situations. Therefore, we typically use $\gamma_3$ (or $\gamma_4$) to parametrize the kinematic path, because throughout the range of folding, the angle $\gamma_3$ (or $\gamma_4$) has a unique value for each configuration. In each mode, the Trimorph unit cell display distinct folding motion, which leads to different mechanical properties of the tessellated metamaterial, such as the sign of Poisson's ratio and shear-normal coupling coefficient. Therefore, we can regard each mode as the fundamental structure of different material phases.

The triclinic bounding box of a Trimorph unit cell is characterized by the three angles: $\eta_1$, $\eta_2$, $\eta_3$, as shown in Fig.~\ref{fig:geom}C. The value of $\eta_1$, the projected angle onto the $xy$-plane, as a function of $\gamma_3$ is plotted in Fig.~\ref{fig:geom}H. For most range of folding, $\eta_1$ stays close to $90 ^ \circ$, especially in the eggbox mode and when $\delta$ is small. Hence, it can be difficult to notice this non-orthogonality on physical models. Similarly, the variation of angles $\eta_2$ and $\eta_3$ are plotted in Fig.~\ref{fig:geom}I. They play important roles when we tessellate the pattern in three dimensions. Unlike $\eta_1$, the other two triclinic angles often deviate significantly from $90 ^ \circ$. 

\section{Results}
\subsection{Geometric Mechanics of the 2D Assembly}
As the system folds across various modes, its properties vary significantly at each folded state. The geometrically dependent mechanics of the Trimorph metamaterial can be captured through the linearized response at an arbitrary folded state. In this work, we mainly discuss two geometry induced mechanical properties: (1) the in-plane stretching and out-of-plane bending responses of the Trimorph metamaterial that are characterized by the corresponding Poisson's ratios; (2) the shear-normal coupling effect that is characterized by the shear coupling coefficient defined as the ratio of the shear strain to a normal strain.

\begin{figure*}[!ht]
	\centering
	\includegraphics[width=0.90\linewidth]{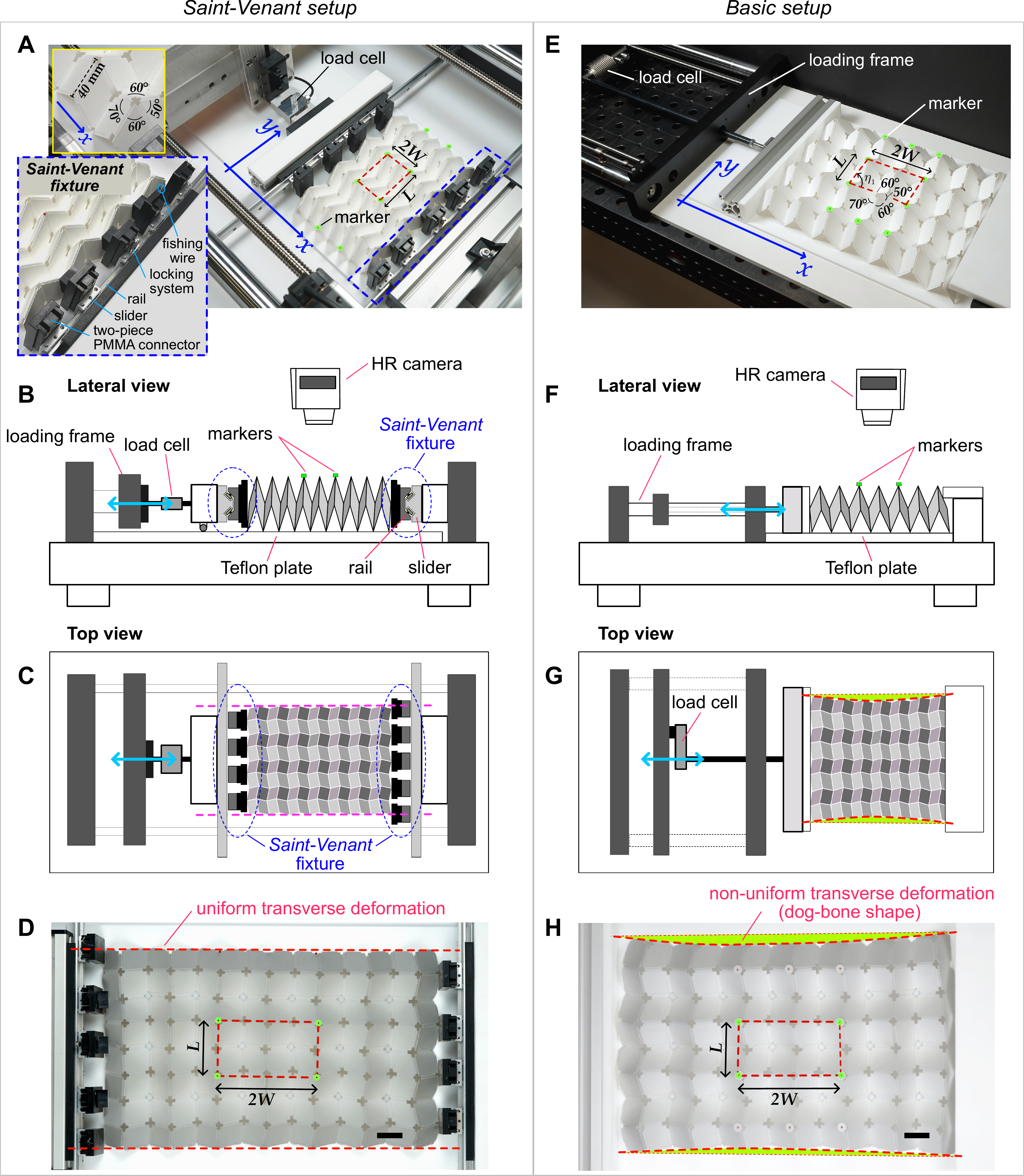}
	\caption{The experimental setup for characterizing the mechanical properties of the Trimorph assembly. (\textbf{A}) Photo and zoom-in details of the \textit{Saint-Venant setup}. (\textbf{B}-\textbf{C}) Design of the \textit{Saint-Venant setup} in lateral and top views, respectively. (\textbf{D}) A snapshot of a sample under testing in the \textit{Saint-Venant setup}. Scale bar:  20mm. (\textbf{E}-\textbf{H}) The photo, design, and sample under testing of the \textit{basic setup}, which is often used in conventional mechanical testing. The non-uniform transverse deformation caused by the \textit{basic setup} reduces the accuracy of the experimental measurements and resulting Poisson's ratio. Scale bar:  20mm.
	}
	\label{fig:mech_setup}
\end{figure*} 

\begin{figure*}[!ht]
	\centering
	\includegraphics[width=0.90\linewidth]{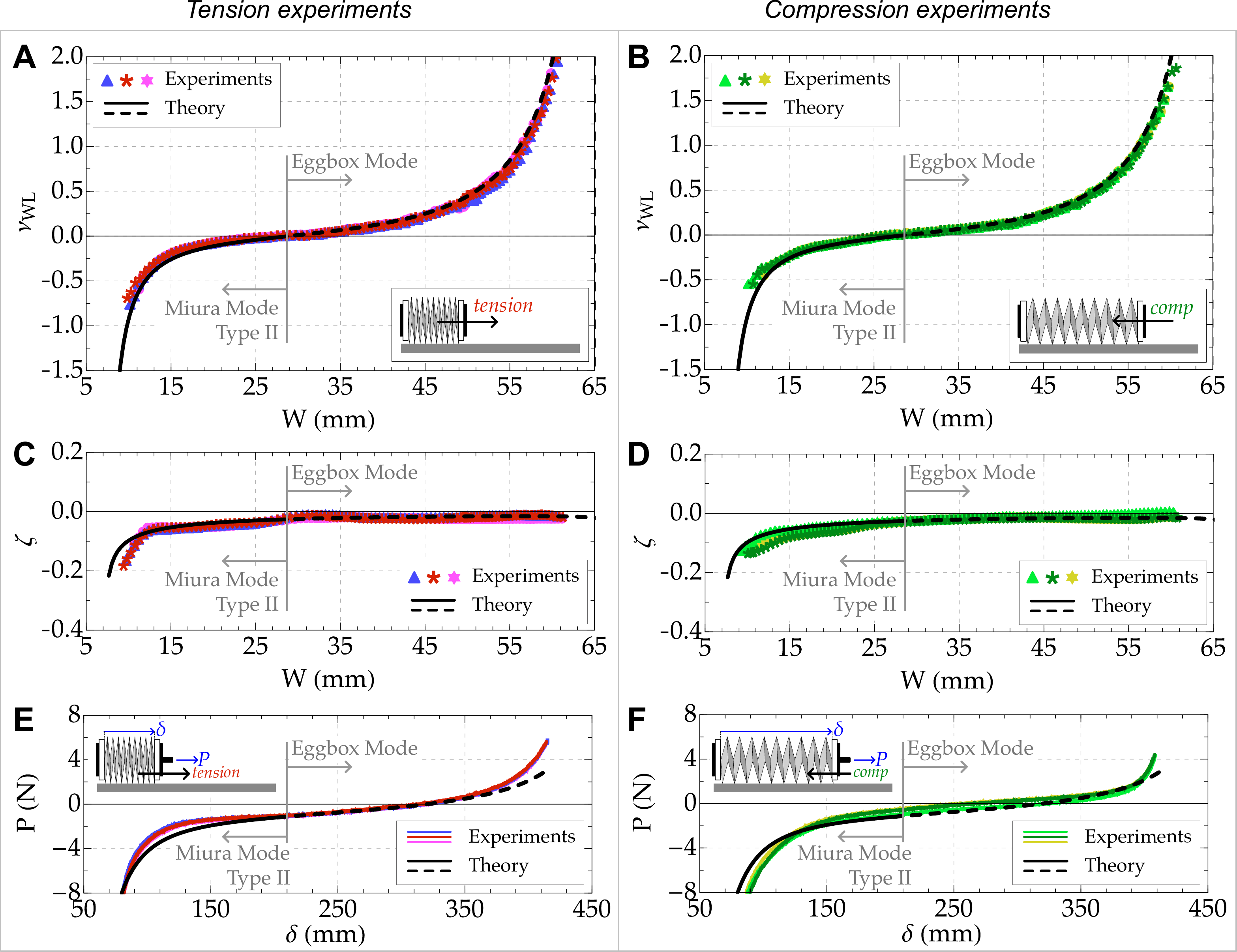}
	\caption{Geometric mechanics of the Trimorph origami-based assembly (2D). (\textbf{A}-\textbf{B}) The lattice Poisson's ratio (LPR) $\nu_{WL}$ vs. average unit cell length $W$, measured in tension and compression tests, respectively. The same sample is tested three times, and the results are shown by different markers. The evaluated coefficients of determination $R^2=0.984\pm0.007$ (tension) and $R^2=0.982\pm0.003$ (compression) indicate an excellent agreement between theory and experiments. (\textbf{C}-\textbf{D}) The shear-normal coupling coefficient $\zeta$ vs. average unit cell length $W$. In this case,  $R^2=0.812\pm0.059$ (tension) and $R^2=0.78\pm0.070$ (compression). (\textbf{E}-\textbf{F}) Nonlinear mechanics behavior through load-displacement diagram. The displacement is defined as the total extension of the entire sample, as illustrated in the insets. Here,  $R^2=0.90\pm0.021$ (tension) and $R^2=0.92\pm0.020$ (compression).}
	\label{fig:mech_data}
\end{figure*} 

We consider a 2D tessellation of the Trimorph unit cell with lattice vectors $\mathbf{W}$ ($O_1O_7$) and $\mathbf{L}$ ($O_7O_9$). Uniform folding of all the unit cells in a tessellation results in in-plane strains of the Trimorph metamaterial. Typically, for isotropic or orthotropic materials, such deformation is characterized by Poisson's ratio, which can be defined as the negative ratio of instantaneous infinitesimal strains along two orthogonal directions \cite{Wei2013,PradeepLiu2019}. For the triclinic Trimorph metamaterial, we define a lattice Poisson's ratio (LPR) to characterize its in-plane deformation, which is defined as the negative ratio of the normal, or extensional, strains along the two lattice directions (i.e. the $\mathbf{L}$ and $\mathbf{W}$ directions). Mathematically, this ratio relates the relative differential change of the angles $\phi$ and $\psi$, and is given by
\begin{equation}  \label{eq:PR1}
	\nu_{WL} =-\frac{\varepsilon_L}{\varepsilon_W}=- \frac{\mathrm{d} L / L}{\mathrm{d} W / W} = -\frac{\tan(\psi/2)}{\tan(\phi/2)}\bigg( \frac{\mathrm{d}\phi}{\mathrm{d}\psi}\bigg) \,. 
\end{equation}
Due to the single degree of freedom nature of the system, we have $\nu_{LW}=1/\nu_{WL}$. As can be noted from Fig.~\ref{fig:geom}D, the slope of the curve (given by the ratio $\mathrm{d}\phi/\mathrm{d}\psi$) is negative for the eggbox mode and positive for the two Miura modes. Therefore, from Eqn.~(\ref{eq:PR1}), the stretching Poisson's ratio is positive for the eggbox mode and negative for the Miura modes. By traversing through the complete kinematic path, the Trimorph pattern takes on values for the lattice Poisson's ratio from the entire set of real numbers, hence displaying reversible auxeticity.
Taking a total differentiation of Eqn.~(\ref{eq:fphipsi}), a closed-form expression for the in-plane stretching Poisson's ratio can be derived as (Supporting Information):
\begin{equation} \label{eq:PR2}
	\nu_{WL} 
	=\frac{\sin^2(\psi/2)}{\sin^2(\phi/2)}\bigg[\frac{\xi_2\cos\phi-2\xi_1+\sin^2\phi\cos\psi}{\xi_2\cos\psi-2\xi_1+\sin^2\psi\cos\phi}\bigg]  \,.
\end{equation}
This expression indicates that the Poisson's ratio for the Trimorph metamaterial is purely a geometric quantity depending only on $\alpha$, $\delta$, and the folded state, independent of the length scale as well as the constituent material of the system. 

Contrasting the in-plane stretching behaviour, out-of-plane bending of the Trimorph metamaterial requires the panels to undergo non-rigid deformation, that simultaneously induces curvatures along the lattice directions. The geometry of the unit cell that corresponds to bending of the system is obtained by imposing quasi-periodicity and frame constraints (Supporting Information). The out-of-plane deformation response is then characterized by the bending-induced lattice Poisson's ratio, which is defined as the negative of the ratio of normal curvatures along the $\mathbf{W}$ and $\mathbf{L}$ directions in the bent configuration. For conventional continuum material, the stretching-induced and bending-induced Poisson's ratio yield same values \cite{timoshenko1982theory}. However, in line with a few studies on origami metamaterials in recent years \cite{Wei2013,nassar2017curvature,PradeepLiu2019}, we also find that the Trimorph metamaterial satisfies the property that the Poisson's ratio in bending and stretching are equal in magnitude but opposite in sign. 

Since the primitive vectors are non-orthogonal for the triclinic metamaterial, then the Poisson effects discussed above deviate from the conventional definition of Poisson's ratios. To address this aspect, we also study the conventional Poisson's ratios along principal directions. Specifically, we define the stretching Poisson's ratio as the negative of the ratio of principal strains, and the bending Poisson's ratio as the negative of the ratio of principal curvatures, which result in evaluations measured along orthogonal directions. We find that the Poisson's ratios defined along the principal directions and the  lattice directions are almost the same. Interestingly, however, the principal Poisson's ratios in bending and stretching are not exactly equal and opposite (Supporting Information).

An interesting biproduct of the non-orthogonal primitive vectors is the shear-normal coupling effect, which relates the shear strain with normal strains. Such effect is useful in some mechanical devices, where the metamaterial is used to transform forces and motions, as a scale-free alternative to traditional mechanisms \cite{frenzel_three-dimensional_2017,lipton_handedness_2018}. A coupling coefficient $\zeta$ is defined to characterize this effect. Denoting $\varepsilon_{WL}$ as the half shear strain induced by normal strain $\varepsilon_W$, we obtain (Supporting Information):
\begin{equation} \label{eq:shearnormal}
	\zeta = -\frac{2 \varepsilon_{WL}}{\varepsilon_W} = 2 \cot{\eta_1} \,,
\end{equation}
with
\begin{equation}
	\cos{\eta_1} = \frac{\cos{\alpha} (\cos{\delta} - 1)}{2 \sin{(\phi/2)} \sin{(\psi/2)}} .
\end{equation}
In the eggbox mode, $\zeta$ stays close to zero, implying a nearly orthotropic symmetry of the Trimorph metamaterial. 

To verify the reversible auxeticity and shear normal coupling of the Trimorph metamaterial, we perform uniaxial tension and compression tests on a physical prototype composed of $7\times4$ unit cells, and tracked the deformations of a sub-region as shown in Fig.~\ref{fig:mech_setup}. For such experiments, we create a new experimental setup, the \textit{Saint-Venant setup} (Fig. \ref{fig:mech_setup}A-D), to alleviate the influence of artificial boundary effect in the traditional setup (Fig. \ref{fig:mech_setup}E-H) that leads to inaccuracy of measurements (Fig. S12). Compared to the traditional setup \cite{Liu2020}, where the sample is clamped by two smooth plates, in the \textit{Saint-Venant setup}, the sample is constrained by a linear slide system that comprises several sliders inserted into a rail, namely the \textit{Saint-Venant fixture}, which allows for a completely free sample deployment. By eliminating the negative impact of the dog-bone shape on the measurement of Poisson's ratio, the \textit{Saint-Venant fixture} notably improves the agreement between experiments and theory, as plotted in Fig. \ref{fig:mech_data}. In summary, the \textit{Saint-Venant setup} permits the testing of relatively small samples, which are reliable in the sense of representing a true periodic system without violating the underlying theoretical hypothesis.

According to Fig.~\ref{fig:mech_data}, the experimentally measured lattice Poisson's ratio (LPR) and coupling coefficient match with the theoretically predicted values, under both tensile and compression testing conditions. We successfully observed the transition of the lattice Poisson's ratio from positive to negative on the changing triclinic frame. To assert the quality of our fabrication method, we also report the load-displacement curve of the sample, which agrees with the theoretically predicted curve based on rigid origami assumption. The derivation of the theoretical curve is elaborated upon in the Supporting Information. To assess the theoretical formulae (Poisson Ratio’s, Shear-Coupling coefficient, Load vs. Displacement) in predicting the observed data, we have computed the mean coefficient of determination $R^2$ and its standard deviation for all the experiments reported in Fig.~\ref{fig:mech_data} (see Supporting Information section B4 for the details). A coefficient of determination $R^2$ equal to 1 indicates the limit case of perfect agreement between theory and experiments. For all cases, the values of $R^2$ indicate a good match between our theory and the experiments, as listed in the caption of Fig. \ref{fig:mech_data}.
Video recording of the experiments performed with the \textit{Saint-Venant} setup and the \textit{basic setup} are provided as Movie S1 and S4, respectively. 

\subsection{Reprogrammable Frustration}
\begin{figure*}[!hp]s
	\centering
	\includegraphics[width=0.9\linewidth]{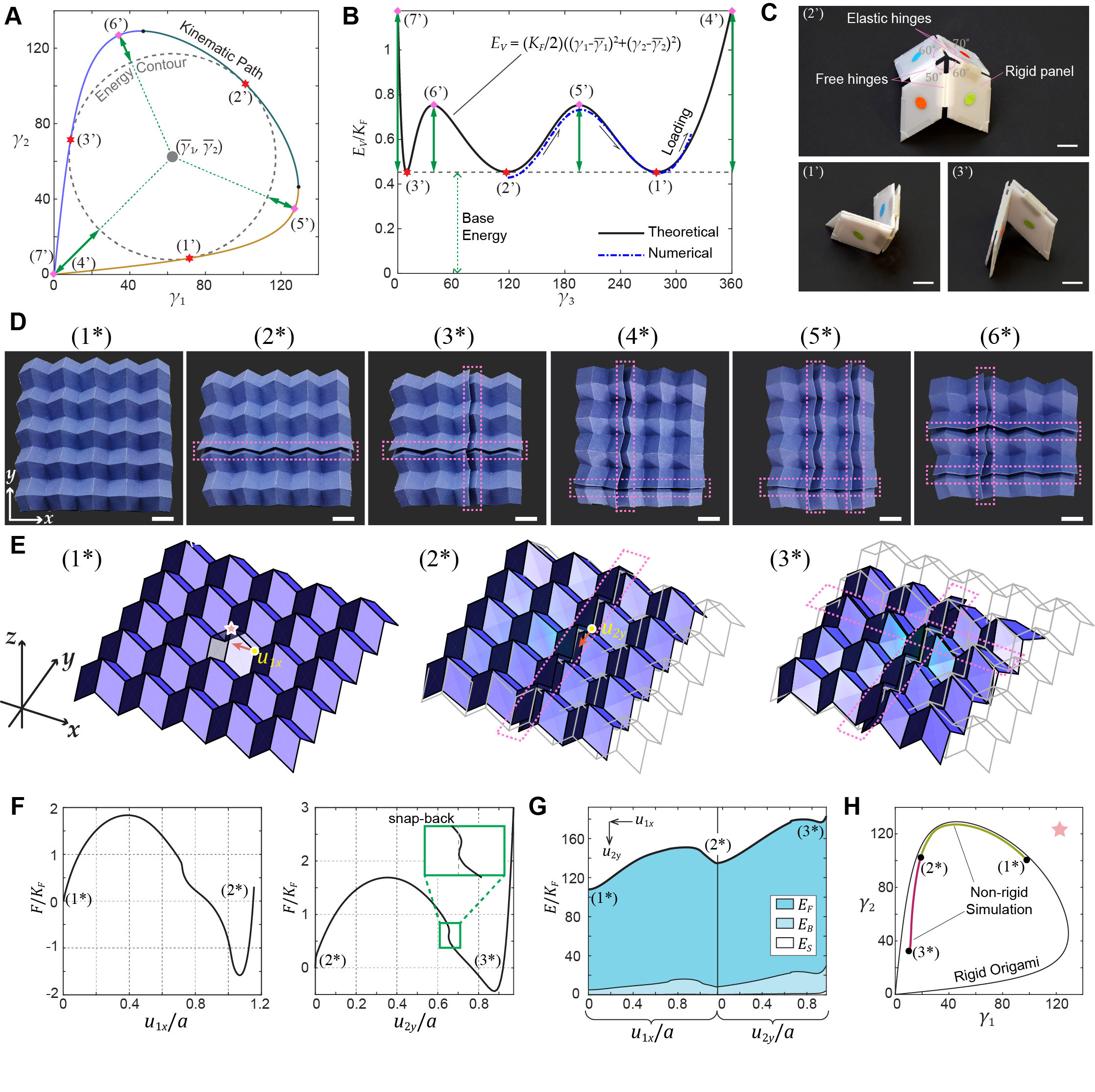}
	\caption{Multistability and reprogrammable frustration of the Trimorph origami. (\textbf{A}) Emergence of tristability. The energy contour has three tangent points with the kinematic path, which indicates three local minima of stored energy. (\textbf{B}) The elastic energy as a function of folding angle $\gamma_3$. The dashed line shows the result from numerical simulation. (\textbf{C}) Photos of the three stable states of a physical model of the Trimorph unit cell. (\textbf{D}) Representative states of a 2D Trimorph assemblage (paper model). The dashed boxes highlight the rows and columns that are ``defected''. Configuration (1*) is the homogeneous state, which marks the ground energy state of the tessellation; Configuration (2*) has one ``line defect''. Configuration (3*) is a frustrated state with two intersecting ``line defects'' and a ``point defect'' at the intersection; Configuration (4*) is another frustrated state with the ``point defect'' at a different location. Configurations (5*) and (6*) are different frustrated states derived from state (4*), each has two ``point defects''. (\textbf{E}) Mechanics setup for numerical simulation of the snapping transitions from state (1*) to (2*), and then to (3*). The dots and arrows show the degrees of freedom that are being traced in the corresponding diagrams. (\textbf{F}) Force vs. displacement curves in the transition from (1*) to (2*), and (2*) to (3*). Notice that the displacement in each diagram is measured on a different degree of freedom, and the inset on the second diagram shows an instance of snap-back. The forces and displacements are normalized. (\textbf{G}) The variation of elastic energy stored in the assemblage during the transition processes. The symbols $E_F$, $E_B$, $E_S$ denote the stored elastic energy caused by folding, bending, and stretching, respectively. (\textbf{H}) The changes of $\gamma_1$ and $\gamma_2$ of the star-marked unit cell in E, i.e. the ``point defect,'' during the transition processes, compared to the kinematic path of a rigid origami Trimorph unit cell. Scale bar: 20mm.}
	\label{fig:tristable}
\end{figure*}
The intrinsic geometry of the Trimorph origami allows for realization of multistability. We model the stored energy $E_V$ of a Trimorph origami unit cell with torsional springs in the folding hinges as:
\begin{equation}
	E_V = \frac{1}{2}\sum_{i=1}^{4} K_{F,i} (\gamma_i - \bar{\gamma_i}) ^ 2\,, 
\end{equation}
where \{$K_{F,i}$\} are the rotational stiffness and \{$\bar{\gamma_i}$\} are the rest angles. This is a theoretical model that follows the rigid origami assumption, which assumes that the origami panels do not deform. When \{$\bar{\gamma_i}$\} do not reside on the rigid folding kinematic path (Fig.~\ref{fig:geom}G), we observe multiple minima of stored energy on the kinematic path \cite{Waitukaitis2015}.  

We design a tristable case for the Trimorph unit cell~\cite{li_theory_2020}, so that there is one local energy minimum in each of the three folding modes. The merit of having each stable state in a different folding mode is that the topological difference between modes leads to significantly different mechanical properties, and thus we can reprogram the properties of the resultant metamaterial by mechanical snapping. To simplify the design and manufacturing, we assign both $O_5 O_6$ ($\gamma_3$) and $O_5 O_8$ ($\gamma_4$) hinges to be free of rotational stiffness (i.e. $K_{F,3} = K_{F,4} = 0$). In addition, we restrict hinges $O_5 O_4$ ($\gamma_1$) and $O_5 O_2$ ($\gamma_2$) to have the same rotational stiffness (i.e. $K_{F,1} = K_{F,2})$, so that the energy contour on the $\gamma_1$ vs. $\gamma_2$ diagram is circular. Normally, such strong simplification will not allow multistability to appear. However, the special folding kinematics of the Trimorph origami makes it possible. Examining the kinematic path of $\gamma_1$ and $\gamma_2$, as shown in Fig.~\ref{fig:tristable}A, we can assign $\bar{\gamma_1}$ and $\bar{\gamma_2}$ at a central point such that the circular energy contour intersects the kinematic path at three tangent points. Due to the symmetry of the kinematic path, $(\bar{\gamma_1}, \bar{\gamma_2})$ must reside on the symmetry axis of the path. Therefore, the two energy minima (1') and (3') become symmetric, each within Miura mode type-I and Miura mode type-II, respectively. The other energy minimum (2') in the eggbox mode happens at the special occasion when $\gamma_1 = \gamma_2$. The change of stored energy in the system is plotted in Fig.~\ref{fig:tristable}B with respect to $\gamma_3$. We note that the tristable unit cell is in a self-stressed state, such that system never rests at a zero-energy state, which can be seen from the non-zero base energy in Fig. \ref{fig:tristable}B.
We can clearly identify three local minima, at the configurations (1'), (2'), and (3'), which are the tangent points in Fig.~\ref{fig:tristable}A. The peaks of energy occur at configurations (4'), (5'), (6'), and (7'), among which, (5') and (6'), (4') and (7'), share the same stored energy ($E_V$). We stress that although configurations (4') and (7') are represented at the same point on the kinematic path, they are not the same as the vertex is flat folded in different orderings. 

To study the transition from one stable configuration to another, we conduct nonlinear structural analyses using the bar-and-hinge model (Movie S3), and consider non-rigid deformations of the panels, i.e. non-rigid origami \cite{Liu2017}. The numerical implementation is detailed in the Methods and Materials. In the numerical simulation, we apply force to push the Trimorph unit cell from the stable configuration (2') to (1'). Because of symmetry, we only perform simulation for the (2') to (1') transition. The stored energy during the snap through process agrees well with the analytical curve, as shown in Fig.~\ref{fig:tristable}B. Overall, the non-rigid numerical model is slightly more compliant than the theoretical rigid origami model. 

To validate our theory, we fabricate physical models (Movie S2). We first make a unit cell comprising of four rigid panels jointed together by four hinges, two free and two elastic, as shown in Fig.~\ref{fig:tristable}C. Details about the fabrication are elaborated upon in the Methods section and the Supporting Information. We observed three stable configurations with the physical model, two Miura modes and one eggbox mode (Fig. \ref{fig:tristable}C and Movie S2). 

When the tristable unit cell is tessellated into a 2D assemblage, the resultant metamaterial displays multiple stable states, as shown in Fig.~\ref{fig:intro}. In the 2D tessellation, each row (a strip of unit cells along the $x$-direction) can transition between the eggbox mode and Miura mode type I, or each column (a strip of unit cells along the $y$-direction) can transition between the eggbox mode and Miura mode type II. This morphing behaviour leads to lines of irregular vertices in the tessellation, resembling a line defect from a crystallographic point of view. The Miura mode changes the primitive vectors of the metamaterial such that the regions in eggbox mode on both sides of a Miura mode strip do not share the same base plane anymore (Fig.~S6). 

This phenomenon exists robustly also for non-rigid origami. We display six out of many possible stable states in Fig. \ref{fig:tristable}D. Assuming rigid origami, as we have shown in the analysis of the unit cell configuration space, the two Miura modes cannot commute without passing through the eggbox mode. Therefore, if one row of unit cells are in Miura mode type-I, and one column of unit cells are in Miura mode type-II, their intersecting unit cell must be within these two modes at the same time, which is forbidden. However, if we consider compliant panels, ``line defects'' in rows and columns would be able to occur simultaneously, as demonstrated by configurations (3*) to (6*). This is possible by having an intersection unit cell that involves not only energy trapped in the folding creases, but also in bent and stretched panels. That is why we need a paper made model to show this scenario, and cannot do the same with the plastic model that is nearly rigid origami. The intersection unit cell is almost crushed and overlaid onto another unit cell, analogous to an interstitial point defect in crystals. 

To understand the formation of the ``point defect,'' we perform nonlinear structural analyses (Movie S3). We first simulate the process of forming a ``line defect'' in a row (x-direction), i.e., transitioning from configuration (1*) to (2*) (Fig.~\ref{fig:tristable}E). Then, based on the configuration (2*), we fold one column to its corresponding Miura mode, i.e. transitioning from configuration (2*) to (3*) (Fig.~\ref{fig:tristable}E). As shown in Fig.~\ref{fig:tristable}F, both processes display snap-through behaviour. Examining the stored energy in the system during the entire process from (1*) to (2*) to (3*), we observe from Fig.~\ref{fig:tristable}G that configuration (3*) stores significantly more energy than (1*) and (2*). This is mainly caused by the non-rigid origami deformation of the intersection unit cell, where the ``point defect'' happens. Fig.~\ref{fig:tristable}H suggests that this unit cell is forced to deviate from its normal kinematic path into a state that significantly deforms the panels, comprising both bending and stretching (Fig.~\ref{fig:tristable}G) deformations. As shown in Fig.~\ref{fig:tristable}D, the frustration can be reprogrammed into different states. 

We perform extra numerical simulations to study the scaling effect of the line and point defects. In addition to the Trimorph pattern consisting of $5 \times 5$ unit cells in Fig. \ref{fig:tristable}, we have added simulations on $3 \times 3$ and $4 \times 4$ patterns. We observe that the line defects exist (without external forces) for all samples sizes, regardless of the number of unit cells. This is owing to the fact that the line defect is a linear combination of natural stable states of the unit cells. However, in our numerical study, the point defect does not appear for $3 \times 3$ and $4 \times 4$ patterns. At the point defect, the unit cell is forced into a highly deformed, frustrated state \cite{sadoc_geometrical_1999} that is not a natural stable state, storing a notable amount of elastic energy. Hence, it can only maintain its local high energy state owing to the kinematic constraints from surrounding unit cells in a tessellation. The effectiveness of such kinematic constraints is a function of the number of unit cells in the corresponding line defects radiating from the point defect. When the constraints from surrounding unit cells are not strong enough, the point defect cannot sustain itself without external forces - it is an unfavourable frustrated state.

To unfold each point defect, the unfolding order must exactly reverse the folding order. For example, if a point defect is formed by first folding a line defect in the $x$-direction and then another in the $y$-direction, this point defect can only be unfolded by first resolving the $y$-direction line defect and then the $x$-direction. This is because the folding order of the unit cell at the point defect becomes different for different forming sequences, as seen from states (1) and (7) of Fig. \ref{fig:geom}D. Due to contact of panels, there is no feasible path to transition from (1) to (7) or vice versa, unless the pattern is unfolded through the entire folding range. In other words, the point defects can lock the pattern if one tries to resolve them in wrong orders. Instead of taking this phenomenon as an disadvantage, we believe that it may become useful for encoding hysteresis information, as mechanical memory for applications in mechanical logic/computing devices \cite{Yasuda2021_natpersp}.

\section{Conclusion}
The Bravais lattices (in general) and the triclinic system (in particular) offer great freedom to create origami-based architected programmable metamaterials. 
Owing to the folding of the origami, the resultant metamaterial can change the six lattice parameters of its triclinic geometry. 
This change of lattice symmetry leads to coupled normal strains and shear strain.
We have demonstrated how origami can be exploited to create anisotropic and inhomogeneous metamaterials, which have properties that are functions of space, orientation, and folding state, resulting in highly tunable responses. 
By tailoring local folding energies, we create a metamaterial that has multiple stable states with distinct configurations, which allows encoding of various phases of matter (see Fig. \ref{fig:table}).
As a result, it transitions from an initially homogeneous tessellation to different inhomogeneous assemblages, as a result of geometric frustration. 
These phenomena are verified experimentally with a standardized manufacturing procedure, showing great potential for engineering applications.

Beyond the elastostatic properties considered in this paper, there are other aspects of this triclinic metamaterial system worth of investigation. For example, material failure behaviour such as fracture pattern, elastodynamic properties such as bandgaps and wave speed, and multi-physical responses such as stimuli responsive actuation, could be addressed in future investigations.




\section{Experimental Section}
\threesubsection{Sample Fabrication}\\
Different types of  unit cells were designed to create i) the multistable 2D tessellation shown in Fig.\ref{fig:intro}(A-G), ii) to carry out the Poisson's ratio experiments reported in Fig. \ref{fig:mech_setup}, and iii) to realize the 3D metamaterials depicted in Fig.\ref{fig:intro}(H,I). The multistable unit cells comprise 4 rigid panels milled with a CNC milling machine from a 2 mm thick Polycarbonate sheet jointed together by 4 hinges, 2 elastic (realized by cutting a silicon rubber solid) and 2 free (milled from a Polypropylene sheet). The unit cells composing the 2D tessellation and the 3D metamaterial were obtained by milling a 1 mm thick Polypropylene sheet. They consist of a single piece of Polypropylene folded from its flat configuration and closed with just one bond. Please see details in Supporting Information. 
The paper model reported in Fig.~S6 is made with Canson Mi-Teintes paper (Canson SAS, France), and we use a Silhouette CAMEO machine (Silhouette America Inc., Utah) to cut the perforated patterns. 

\threesubsection{Mechanical Characterization}\\
The reversible auxeticity of the 2D tessellation was verified using the experimental setup reported in Fig.~\ref{fig:mech_setup}A. The compression/tensile experiments were performed by imposing a constant speed of 1.5 mm/s at one end of the sample with a $\mu$-strain testing machine. Four black markers (1 mm in diameter), located along the sides of a rectangular region in the middle of the sample (Fig.~\ref{fig:mech_setup}A), were used to determine the Poisson’s ratio of the tessellation. The displacements of each marker were determined by a post-processing analysis of the records of the experiments. The compression/tension experiments were performed by imposing a constant speed of 1.5 mm/s at one end of the sample with the testing machine. Such a speed was carefully chosen, combining the need to ensure the quasi-static condition and the requirement to reduce the stick and slip phenomena between the sample and the testing Teflon platform. In particular, a higher speed would have effected the measurements with spurious inertia contribution.
Please see details in the Supporting Information. 

\threesubsection{Numerical Simulations}\\
The numerical simulations are performed using the MERLIN software \cite{Liu2018}. The software implements the bar-and-hinge model for discretization of origami structures. We adopt the N5B8 model \cite{Filipov2017}, which discretizes each quadrilateral panel into four triangles, and represent the origami behavior by bars and torsional springs, which captures three essential deformation modes: folding, panel bending, and stretching. The elastic energy stored in the bars and hinges compose the system elastic energy. The quasi-static response of the structure is then obtained by finding the stationary states of the system energy, using the Modified Generalized Displacement Control Method. It has been shown by experiments that the accuracy of the bar-and-hinge model is surprisingly good. In this work, we take the folding stiffness parameter $K_F$ to be 1/10 of the bending stiffness parameter $K_B$, which represents a typical non-rigid origami. Other input information such as the detailed boundary conditions for the simulations in this paper can be read from the input files to the MERLIN software (version 2), shared in the Supporting Information.

\medskip
\textbf{Supporting Information} \par 
Supporting Information is available from the Wiley Online Library or from the author.

\medskip
\textbf{Acknowledgements} \par 
The authors thank the support from the US National Science Foundation (NSF) through grant no.1538830. K.L acknowledges the support from Peking University College of Engineering. P.P.P. acknowledges the support from the Indian Institute of Technology Madras through the seed grant and the Science \& Engineering Research Board (SERB) of the Department of Science \& Technology, Government of India, through award SRG/2019/000999. D.M. is supported by the European Commission under the H2020 FET Open (“Boheme”) grant No. 863179 and by the ERC-ADG-2021-101052956-BEYOND. T.T. is supported by Japan Science and Technology Agency PRESTO JPMJPR1927.

\medskip

%
\bibliographystyle{MSP}
\bibliography{Trimorph}

\end{document}